\documentclass[12pt]{article}
\usepackage{amsmath}
\numberwithin{equation}{section}
\newcommand{\ma}{\mathcal}
\begin{document}
\title{Simplification of additivity conjecture in quantum information theory}
\author{Motohisa Fukuda 
\footnote{Email: m.fukuda@statslab.cam.ac.uk}\\
Statistical Laboratory,\\
Centre for Mathematical Sciences,\\
University of Cambridge} 

\maketitle
 
\abstract{
We simplify some conjectures in quantum information theory;
the additivity of minimal output entropy,
the multiplicativity of maximal output $p$-norm and
the superadditivity of convex closure of output entropy.
We construct a unital channel for a given channel so that
they share the above additivity properties;
we can reduce the conjectures for all channels 
to those for unital channels. 
}

\section{Introduction}

It is natural to measure the noisiness of a
(quantum) channel by the minimal output entropy (MOE);
how close can the output be to a pure state 
in terms of the von Neuman entropy.
It is then important to ask if
a tensor product of two channels can ever be less noisy in the sense
that some entangled input can have its output closer to a pure state
than the product of the optimal inputs of the two channels.    
This leads to the additivity conjecture of MOE.
Actually, the additivity of MOE 
has been shown \cite{AB04},\cite{Sho03},\cite{Pom03} to be
equivalent globally to several other fundamental
conjectures in quantum information theory;
the additivity of Holevo capacity,
the additivity of entanglement of formation and
the strong superadditivity of entanglement of formation.
In this papar we write not only about 
the additivity of MOE 
but also about two other conjectures;
the multiplicativity of maximal output $p$-norm,
which measures how close can the output be to a pure state 
in terms of $p$-norm,
and the superadditivity of convex closure of output entropy.
Note that
the multiplicativity of maximal output $p$-norm implies the additivity of MOE \cite{Hol02},
and the superadditivity of convex closure of output entropy
the additivity of entanglement of formation \cite{MSW04},\cite{Sho03}.
In this sense, these two conjectures are stronger than
the above four equivalent conjectures.

Curiously, most channels for which 
the additivity of MOE has been proven are unital:
unital qubit channels \cite{Kin02},
the depolarizing channel \cite{FH02},\cite{Kin03},\cite{Amo04}, 
the Werner-Holevo channel \cite{MY04},\cite{DHS04},\cite{AF04},
the transpose depolarizing channel \cite{FHMV04},\cite{DHS04a},
and some asymmetric unital channels \cite{DR05}.
By contrast,
non-unital channels have been extremely resistant
to proofs.
Of course there are some proofs on non-unital channels;
entanglement-breaking channels \cite{Sho02},\cite{Kin02a},
a modification of the Werner-Holevo channel \cite{WE05}
and diagonal channels \cite{Kin05}.
However in the paper \cite{Fuk05} 
we had to extend the result on the depolarizing channel,
which is unital, to non-unital ones.

In this paper 
we simplify the three conjectures;
the additivity of MOE,
the multiplicativiy of maximal output $p$-norm and
the superadditivity of convex closure of output entropy.
We show that 
proving these conjectures for a product of any two unital channels is enough 
by using some unital extension of channels.
This is significant because having proven this
we don't have to consider non-unital channels
as long as these conjectures are concerned.

Let us give some basic definitions.
A (quantum) state is represented as a positive 
semidefinite operator $\rho$ of trace one in a Hilbert 
space $\mathcal{H}$; this is called a 
density operator.
We denote the sets of all bounded operators and 
all density operators in $\mathcal{H}$ by
${\mathcal B}(\mathcal{H})$ and ${\mathcal D}(\mathcal{H})$ respectively.
A (quantum) channel $\Phi$ from $\ma{H}_1$ to $\ma{H}_2$ is 
a completely positive (CP) 
trace-preserving (TP) map (CPTP map) 
from $\mathcal{B}(\mathcal{H}_1)$ to $\mathcal{B}(\mathcal{H}_2)$. 
A channel $\Phi$ is called bistochastic if
\begin{align*}
\Phi(\bar{I}_{\ma{H}_1})=\bar{I}_{\ma{H}_2}.
\end{align*}
Here $\bar{I}_{\ma{H}_1}=I_{\ma{H}_1}/{\rm dim}\ma{H}_1$,
called the normalised identity,
where $I_{\ma{H}_1}$ is the identity operator in $\ma{H}_1$
($\bar{I}_{\ma{H}_2}$ is similarly defined).
When $\ma{H}_1 = \ma{H}_2$ bistochastic channels are called unital channels.

The MOE of a channel $\Phi$ is defined as
\begin{align}
S_{\rm min}(\Phi):=\inf_{\rho \in {\mathcal D}(\mathcal{H}) }S(\Phi(\rho)),
\end{align}
where $S$ is the von Neumann entropy:
$S(\rho)=-{\rm tr}[\rho\log\rho]$.
The additivity conjecture of MOE \cite{KR01} is that
\begin{align} 
S_{\rm min}(\Phi \otimes \Omega)= S_{\min}(\Phi)+S_{\min}(\Omega)
\end{align}
for any channels $\Phi$ and $\Omega$.
Note that the bound $S_{\min}(\Phi \otimes \Omega) \leq  S_{\min}(\Phi)+S_{\min}(\Omega)$ 
is straightforward.

The maximal output $p$-norm of a channel $\Phi $ is defined as
\begin{align}
\nu_p(\Phi):=\sup_{\rho \in {\mathcal D}(\mathcal{H})}\|\Phi(\rho)\|_p,
\end{align}
where $\|\;\;\|_p$ is the Schatten $p$-norm: $\|\rho\|_p=({\rm{tr}}|
\rho|^p)^{\frac{1}{p}}$.
The multiplicativity conjecture of maximal output $p$-norm is that
\begin{align} \label{mult}  
\nu_p(\Phi \otimes \Omega)= \nu_p(\Phi)\nu_p(\Omega)
\end{align}
for any channels $\Phi$ and $\Omega$, and any $p \in [1,2]$.
The multiplicativity was conjectured to be true for $p\in[1,\infty]$
before a counterexample was found \cite{WH02}.
Note that the bound 
$\nu_p(\Phi \otimes \Omega) \geq  \nu_p(\Phi)\nu_p(\Omega)$ 
is straightforward.

The convex closure of output entropy of 
a channel $\Phi$ 
is 
\begin{align} 
H_{\Phi}(\rho)
={\rm min} \left\{ \sum_i p_i S(\rho_i):
\sum_i p_i \rho_i = \rho, \sum_i p_i =1, p_i\geq 0
\right\}
\end{align}
for a state $\rho$.
The superadditivity conjecture of convex closure of output entropy is that
\begin{align}
H_{\Phi \otimes \Omega}(\rho)
\geq H_{\Phi}(\rho_{\ma{H}}) + H_{\Omega}(\rho_{\ma{K}}).
\end{align}
for any channels $\Phi$ and $\Omega$, and 
any state $\rho \in \ma{D}(\ma{H} \otimes \ma{K})$.
Here 
the input spaces of $\Phi$ and $\Omega$ are 
$\ma{H}$ and $\ma{K}$ respectively, and
$\rho_{\ma{H}}={\rm tr}_{\ma{K}}[\rho]$ and 
$\rho_{\ma{K}}={\rm tr}_{\ma{H}}[\rho]$.

We introduce the Weyl operators to be used later.
Given a Hilbert space $\mathcal{H}$ of dimension $d$
let us choose an orthonormal basis $\{e_{k};k=0,\dots,d-1\}$.
Consider the additive cyclic group $\mathbf{Z}_{d}$ and define an irreducible
projective unitary representation of the group 
$ Z=\mathbf{Z}_{d}\oplus \mathbf{Z}_{d} $ 
in $\mathcal{H}$ as
\begin{align*}
z=(x,y)\mapsto W_{z}=U^{x}V^{y},
\end{align*}
where $x,y\in $ $\mathbf{Z}_{d},$ and $U$ and $V$ are the unitary operators
such that
\begin{align*}
U|e_{k}\rangle=|e_{k+1(\mathrm{mod}d)}\rangle,\qquad V|e_{k}\rangle=\exp \left(
\frac{2\pi \mathrm{i} k}{d}\right) |e_{k}\rangle.
\end{align*}
Then we have
\begin{align} \label{complete noise}
\sum_z W_z \rho W_z^\ast=d^2 \bar{I}_{\ma{H}},
\qquad 
\forall \rho \in \ma{D}(\ma{H}).
\end{align}

\section{Result}

We were inspired by the Shor's paper \cite{Sho03} 
to find the following theorem:
\\
{\bf Theorem 1.}
{\sl
Take a channel $\Omega$.
\\
1) The additivity of MOE of
$\Phi_{\rm b} \otimes\Omega$ for any bistochastic channel $\Phi_{\rm b}$
would imply that of $\Phi \otimes\Omega$ for any channel $\Phi$.
\\
2) Fix $p \in [1,\infty]$. 
The multiplicativity of maximal output $p$-norm of
$\Phi_{\rm b} \otimes\Omega$ for any bistochastic channel $\Phi_{\rm b}$
would imply
that of $\Phi \otimes\Omega$ for any channel $\Phi$.
\\
3) The superadditivity of convex closure of output entropy of
$\Phi_{\rm b} \otimes\Omega$ for any bistochastic channel $\Phi_{\rm b}$
would imply that of $\Phi \otimes\Omega$ for any channel $\Phi$.
}

{\bf Proof.}
1)
Suppose we have a channel 
\begin{align*}
\Phi : \ma{B}( \ma{H}_1) &\longrightarrow  \ma{B}(\ma{H}_2).
\end{align*} 
Let $d= {\rm dim}\ma{H}_2$.
Then we construct a new channel $\Phi^\prime$: 
\begin{align*}
\Phi^\prime: \ma{B}({\bf C}^{d^2}\otimes \ma{H}_1)  
&\longrightarrow \ma{B}(\ma{H}_2) \\
\tilde{\rho} &\longmapsto 
\sum_{z} W_z \Phi (E_z \tilde{\rho} E_z^\ast ) W_z^\ast.
\end{align*}
Here 
$W_z$ are the Weyl operators in $\ma{H}_2$ and
$E_z = (\langle z | \otimes I_{\ma{H}_1})$, 
where $\{|z\rangle\}$ forms the standard basis for ${\bf C}^{d^2}$.
Note that this channel is bistochastic by (\ref{complete noise}).

First, we show
\begin{align}\label{unital_extension}
S_{\rm min}(\Phi^\prime \otimes \Omega) 
\leq S_{\rm min}(\Phi \otimes \Omega),
\end{align}
for any channel $\Omega$.
Suppose $\Omega$ is a channel such that 
$\Omega : \ma{B}(\ma{K}_1)\rightarrow \ma{B}(\ma{K}_2)$.
Then 
\begin{align*}
(\Phi^\prime &\otimes \Omega)(|(0,0)\rangle\langle (0,0)|\otimes \rho)\\
&=({\bf 1}_{\ma{H}_2}\otimes\Omega)
((\Phi^\prime \otimes {\bf 1}_{\ma{K}_1})(|(0,0)\rangle\langle (0,0)| \otimes \rho))\\
&=({\bf 1}_{\ma{H}_2}\otimes\Omega )
(( \Phi  \otimes {\bf 1}_{\ma{K}_1} )(\rho))\\
&=(\Phi \otimes \Omega)(\rho),
\end{align*}
for any $\rho \in \ma{D}(\ma{H}_1 \otimes \ma{K}_1)$.
Here $(0,0)\in \mathbf{Z}_{d}\oplus \mathbf{Z}_{d}$
and $W_{(0,0)}=I_{\ma{H}_2}$.

Next, we show the converse;
\begin{align}\label{unital_extension_reverse}
S_{\rm min}(\Phi^\prime \otimes \Omega)
\geq S_{\min}(\Phi \otimes \Omega).
\end{align}
for any channel $\Omega$.
Take $\hat{\rho} \in \ma{D}({\bf C}^{d^2} \otimes \ma{H}_1\otimes \ma{K}_1) $.
Let $\rho_z = (E_z \otimes I_{\ma{K}_1})\hat{\rho}( E_z \otimes I_{\ma{K}_1})$,
$c_z = {\rm tr} \rho_z$ and $\bar{\rho}_z = \rho_z / c_z$
for $z \in Z=\mathbf{Z}_{d}\oplus \mathbf{Z}_{d}$. 
Note that $\hat{\rho}$ can be written in the matrix form:
\begin{align*}
\hat{\rho}=
\begin{pmatrix}
\rho_{(0,0)} & \ldots & \ast\\
\vdots & \ddots & \vdots \\
\ast & \ldots & \rho_{(d-1,d-1)}
\end{pmatrix}.
\end{align*}
This is a $d^2 \times d^2$ block matrix, 
where each block is an element in $\ma{B}(\ma{H}_1 \otimes \ma{K}_1)$.
Then, by concavity of the von Neumann entropy,
\begin{align*}
S\left((\Phi^\prime \otimes \Omega )(\hat{\rho})\right)
&= S\left(\sum_z (W_z \otimes I_{\ma{K}_2})( (\Phi \otimes \Omega)( \rho_z)) 
(W_z^\ast \otimes I_{\ma{K}_2})\right)\\
&\geq \sum_z c_z S((\Phi\otimes \Omega)(\bar{\rho}_z))\\
&\geq S_{\rm min}(\Phi \otimes \Omega).
\end{align*}

Finally, since we assumed the additivity for a product
of any bistochastic channel and $\Omega$
(\ref{unital_extension}) and (\ref{unital_extension_reverse}) show
\begin{align*}
S_{\rm min}(\Phi \otimes \Omega)
= S_{\rm min}(\Phi^\prime \otimes \Omega)
=S_{\rm min}(\Phi^\prime)+S_{\rm min}(\Omega)
=S_{\rm min}(\Phi)+S_{\rm min}(\Omega). 
\end{align*}

2)
A proof on the multiplicativity can be obtained in a similar way as above.

3)
As in the proof 1) take any channel $\Omega$ to have the following result:
\begin{align*}
H_{\Phi^\prime \otimes \Omega}(|(0,0) \rangle\langle (0,0)| \otimes \rho)
&= {\rm min} \sum_i p_i 
S((\Phi^\prime \otimes \Omega)(|(0,0) \rangle\langle (0,0)| \otimes \rho_i))\\
&= {\rm min} \sum_i p_i 
S(( \Phi \otimes \Omega)(\rho_i)) \\
&= H_{\Phi \otimes \Omega}(\rho)
\qquad \forall \rho \in \ma{D}(\ma{H}_1 \otimes \ma{K}_1).
\end{align*}
To see the first equality note that
\begin{align*}
|(0,0)\rangle\langle (0,0)| \otimes\rho=\sum_ip_i \hat{\rho}_i
\qquad\Rightarrow\qquad
\hat{\rho}_i = | (0,0)\rangle\langle (0,0)| \otimes \rho_i
\qquad \forall i.  
\end{align*}
Here, $\rho_i \in \ma{D}(\ma{H}_1 \otimes \ma{K}_1)$.
By the assumption we have
\begin{align*}
H_{\Phi \otimes \Omega}(\rho) 
&= H_{\Phi^\prime \otimes \Omega}(|(0,0) \rangle\langle (0,0)| \otimes \rho)\\
&\geq H_{\Phi^\prime}(|(0,0) \rangle\langle (0,0)| \otimes \rho_{\ma{H}_1})
+H_{\Omega}(\rho_{\ma{K}_1}) \\
&= H_{\Phi}(\rho_{\ma{H}_1})
+H_{\Omega}(\rho_{\ma{K}_1})
\end{align*}
QED

{\bf Remark.}
The first part of proof 1) shows that 
$\Phi^\prime$ is a bistochastic extension of $\Phi$.
In the following corollary we form a unital extension of $\Phi$.

{\bf Corollary 2.}
{\sl In each case of theorem 1,
the assumption would be implied by
proving the conjecture on $\Phi_{u} \otimes \Omega$ 
for all unital channels $\Phi_{u}$.}

{\bf Proof.}
Consider a unital channel:
\begin{align*}
\Phi^{\prime\prime}: \ma{B}({\bf C}^{d^2}\otimes \ma{H}_1 )
&\longrightarrow \ma{B}({\bf C}^{cd}\otimes \ma{H}_2) \\
\tilde{\rho}
&\longmapsto \bar{I}_{{\bf C}^{cd}} \otimes \Phi^\prime(\tilde{\rho}).
\end{align*}
Here $c$ is the dimension of $\ma{H}_1$.
Then it is not difficult to see
\begin{align*}
S((\Phi^{\prime\prime} \otimes \Omega) (\hat{\rho}))
&= \log cd + S((\Phi^\prime \otimes \Omega)(\hat{\rho})) \\
\| (\Phi^{\prime\prime} \otimes \Omega) (\hat{\rho})\|_p 
&= (cd)^{\frac{1-p}{p}}
\|(\Phi^{\prime} \otimes \Omega )(\hat{\rho})\|_p.
\end{align*}
for any channel $\Omega$ and 
any state $\hat{\rho} \in \ma{D}({\bf C}^{d^2}\otimes\ma{H}_1\otimes \ma{K}_1)$.
The results follow obviously.
QED

{\bf Corollary 3.}
{\sl
The following statements are true.
\\
1) The additivity of MOE of
$\Phi_{\rm u} \otimes\Omega_{\rm u}$ for any unital channels 
$\Phi_{\rm u}$ and $\Omega_{\rm u}$
would imply that of $\Phi \otimes\Omega$ 
for any channels $\Phi$ and $\Omega$. 
\\
2) Fix $p \in [1,\infty]$. 
The multiplicativity of maximal output $p$-norm of
$\Phi_{\rm u} \otimes\Omega_{\rm u}$ for any unital channels 
$\Phi_{\rm u}$ and $\Omega_{\rm u}$
would imply
that of $\Phi \otimes\Omega$ for any channels $\Phi$ and $\Omega$.
\\
3) The superadditivity of convex closure of output entropy of
$\Phi_{\rm u} \otimes\Omega_{\rm u}$ for any unital channels 
$\Phi_{\rm u}$ and $\Omega_{\rm u}$
would imply that of $\Phi \otimes\Omega$ 
for any channels $\Phi$ and $\Omega$.
}

{\bf Remark.}
In order to prove the theorem 
we generalized the channel extension which
Shor used \cite{Sho03} to prove that
the additivity of Holevo capacity implies 
the additivity of MOE.
Reading his paper in view of our paper
one can notice that
proving the additivity of Holevo capacity for all unital channels
would imply the additivity for all channels.

\section{Conclusion} 
By using the results in this paper
we can focus on unital channels 
to prove the additivity of MOE,
the multiplicativity of maximal output $p$-norm,
the superadditivity of convex closure of output entropy
and the additivity of Holevo capacity,
or to find a counterexample.
\\\\

\begin{center}
{\bf Acknowledgement}
\end{center}
I would like to thank my supervisor Yuri Suhov
for suggesting the problem,
constant encouragement and numerous discussions.
I also would like to thank Alexander Holevo 
for giving useful comments and 
especially pointing out that the theorem also works for 
the superadditivity of convex closure of output entropy.
M. B. Ruskai is thanked for making useful comments.

\end{document}